\documentclass[aps,twocolumn,showpacs,floatfix,superscriptaddress]{revtex4}
\usepackage{graphicx,amsmath,amssymb,times}
\usepackage{txfonts}
\usepackage{psfrag}
\usepackage[usenames]{color}
\usepackage[colorlinks,urlcolor=blue,citecolor=blue,linkcolor=blue]{hyperref}
\usepackage{subfigure}
\usepackage{grffile}

\begin{document}
\newcommand{\p}[1]{\textrm{#1}}
\newcommand{\mb}[1]{\mathbf{#1}}
\newcommand{\reffig}[1]{\mbox{Fig.~\ref{#1}}}
\newcommand{\refeq}[1]{\mbox{Eq.~(\ref{#1})}}
\newcommand{\rbcs}{$^{87}$Rb$^{133}$Cs}
\newcommand{\krb}{$^{40}$K$^{87}$Rb}
\newcommand{\note}[1]{\textcolor{red}{[\textrm{#1}]}}

  \title{Resonant control of polar molecules in an optical lattice}

  \author{Thomas M. Hanna}
  \affiliation{Joint Quantum Institute,
  NIST and University of Maryland,
  100 Bureau Drive, Stop 8423,
  Gaithersburg, MD 20899-8423, USA}  
  \author{Eite Tiesinga}
  \affiliation{Joint Quantum Institute,
  NIST and University of Maryland,
  100 Bureau Drive, Stop 8423,
  Gaithersburg, MD 20899-8423, USA}
  \author{William F. Mitchell}
  \affiliation{Applied and Computational Mathematics Division, National Institute of Standards and Technology, 100 Bureau Drive Stop 8910, Gaithersburg, Maryland 20899-8910, USA}
  \author{Paul S. Julienne}
  \affiliation{Joint Quantum Institute,
  NIST and University of Maryland,
  100 Bureau Drive, Stop 8423,
  Gaithersburg, MD 20899-8423, USA}
    
  \begin{abstract}
We study the resonant control of two nonreactive polar molecules in an optical lattice site, focussing on the example of RbCs. 
Collisional control can be achieved by tuning bound states of the intermolecular dipolar potential, by varying the applied electric field or trap frequency.
We consider a wide range of electric fields and trapping geometries, showing that a three-dimensional optical lattice allows for significantly wider avoided crossings than free space or quasi-two dimensional geometries.
Furthermore, we find that dipolar confinement induced resonances can be created with reasonable trapping frequencies and electric fields, and have widths that will enable useful control in forthcoming experiments.
    \end{abstract}

  \date{\today}
\maketitle

Ultracold gases of polar molecules are of interest for their long-range dipolar interactions, which give them unique applications in areas such as many-body phases~\cite{baranov08}, quantum information~\cite{demille02}, and precision measurement~\cite{sanders67, demille00, hudson02}. 
Cold gases of LiCs~\cite{dieglmayer08} and RbCs~\cite{sage05} have been formed with temperatures $T \lesssim 1\,$mK, and work continues to produce degenerate gases~\cite{lercher11, cho11}.
Since the creation of a near-degenerate gas of \krb\ with $T \lesssim 1\,\mu$K~\cite{ni08science}, a number of studies have been done of its collision properties~\cite{ni10, ospelkaus10prl, ospelkaus10science, demiranda11}.
KRb has an exothermic reaction producing $\mathrm{K}_2 + \mathrm{Rb}_2$, which occurs with almost unit probability when two molecules are sufficiently close.
This allows a simple description of the collision properties in terms of universal physics~\cite{quemener10, quemener10b, micheli10, idziaszek10, gao10, kotochigova10, quemener11, julienne11}.
Such techniques should also apply to other species with reactive collisions, and to the quenching of any vibrationally excited molecule.
For many studies it is therefore desirable to keep molecules separated, for example by confining the gas in a three-dimensional (`3D') lattice~\cite{chotia11} or in a quasi-two dimensional (`2D') geometry with the molecules polarized perpendicular to the plane (`side-by-side')~\cite{demiranda11, quemener10, quemener10b, micheli10, ticknor10, dincao11}. 
This reduces the likelihood of molecules approaching each other along the attractive `head-to-tail' path which is available in 3D.

In contrast to reactive molecules such as KRb, ground state NaK, NaRb, NaCs, KCs and RbCs are not reactive, and so are available for experiments on longer timescales and at higher densities where control of elastic collisions is useful.
The long range dipole-dipole interaction between two molecules produces an anisotropic potential which is capable of supporting bound states~\cite{kanjilal08}.
Tuning these bound states around a collision threshold with an electric field allows resonant control of the interactions~\cite{ticknor05, roudnev09, idziaszek10}, in analogy to the magnetic and optical control that has been so useful for neutral atoms~\cite{chin_review}.
Because three-body recombination can still occur~\cite{ticknor10prl}, isolating a pair of molecules in an optical lattice site provides an ideal, loss-free environment for studying the two-body energy spectrum.
Such a scenario is analogous to several experiments performed on atom pairs~\cite{syassen07, ospelkaus06, thalhammer06, volz06}.
Optical lattices have also been used to tune atomic collisions through confinement induced resonances~\cite{olshanii98, petrov00, sala11, haller10, froehlich11} (CIRs), which depend on the scattering length being comparable to the characteristic length of the confinement.
With the interactions of polar molecules having an even longer range, it is reasonable to anticipate easy creation of a CIR.

In this paper we study the states and control possibilities of two polar molecules isolated in an optical lattice site, focussing on the specific example of RbCs.
We examine in detail the effects of tuning the lattice parameters and electric field.
We show that the optical lattice can be used to increase the resonance width past what is possible in free space or 2D geometries.
We compare the eigenenergies obtained for a quasi-2D lattice site to scattering calculations for a system with confinement in only one direction, which accurately reproduce the avoided crossings and show their utility for resonant control.
Our studies show that tuning the confinement has a significant effect on the collisional and bound state properties of the pair of molecules, allowing the creation of useful CIRs.

\begin{center}
\begin{table}[tb]
\begin{tabular}{|c | c | c |}
\hline
Quantity & Definition & Value ($a_0$) \\ \hline
Mean scattering length 	& $ \bar{a} =  \frac{2\pi}{[\Gamma(1/4)]^2} (2m_\p{r} C_6/\hbar^2)^{1/4} $ & 233.5 \\
Confinement length & $\ell_\mathrm{ho} = \sqrt{\hbar/(2m_\p{r} \omega)}$	 & \\
$\omega/2\pi = 1$\,kHz				&				& 5728 \\
$\omega/2\pi = 50$\,kHz				&				& 810.5 \\
Dipole length			& $a_\mathrm{\mu} = m_\p{r} \mu^2 / \hbar^2 $						& $3.1 \times 10^{4}$ \\ \hline 
\end{tabular}
\caption{Characteristic length scales for the interaction of ultracold molecules in an optical lattice, with values given for RbCs in parameter regimes used in the present work. For the van der Waals coefficient, we use $C_6 = 142129 E_h a_0^6$~\cite{kotochigova10}, where $E_h = 4.3597 \times 10^{-18}$\,J is the Hartree energy and $a_0 = 52.918$\,pm is the Bohr radius. We give the confinement length for optical lattice sites with frequencies $\omega/2\pi = 1\,$kHz and 50\,kHz.
The dipole length is given for RbCs molecules with a dipole moment of $\mu = 1.0\,$D, where $\p{D} = 0.39343 ea_0 = 3.336 \times 10^{-30}$\,Cm is the Debye and $e$ is the charge of an electron. Here, $m_\p{r}$ is the reduced mass.
}
\label{table:length_scales}
\end{table}
\end{center}

We consider two ground state $^{87}$Rb$^{133}$Cs molecules in a cylindrically symmetric optical lattice site.
In Table~\ref{table:length_scales} we list length scales relevant to ultracold molecular collisions and give representative values for the parameter regimes used in this work.
Typical van der Waals coefficients for polar molecules are of order $10^{5}\,E_h a_0^6$ -- $10^{7}\,E_h a_0^6$, much larger than those for pairs of alkali atoms ($10^3 \,E_h a_0^6$ -- $10^{4}\,E_h a_0^6$). 
However, the mean scattering length scales as $\bar{a} \propto C_6^{1/4}$, giving a similar characteristic length to the van der Waals part of the potential.
We take the dipole moment $\mu = \langle \hat{\mu} \rangle_\p{z}$ to be the expectation value of the electric dipole operator $\hat{\mu}$ for the molecular ground state in the electric field direction.
We calculate this electric-field dependent quantity according to the method of Ref.~\cite{bohn_chapter}.
The dipole length, tunable with an electric field, is typically the largest length scale in the problem. 
For a trapping frequency $\omega/2\pi = 50\,$kHz, $a_\mu = \ell_\p{ho}$ for $\mu = 0.16\,$D.
We note that the use of a strong dipole moment takes us beyond the region of validity of pseudopotential approaches such as those of Refs.~\cite{kanjilal07, derevianko03}.

The molecules are assumed to be rigid rotors, aligned in the axial direction by an applied electric field.
We approximate the lattice site with a harmonic trap and consider only the relative motion of the molecules. 
The combined interaction and trapping potential is given by
\begin{align}
V(\rho, z) &= \frac{\mu^{2}}{r^3} \left(1 - \frac{3z^2}{r^2} \right) + \frac{C_{12}}{r^{12}} - \frac{C_6}{r^6} + 
\frac{\hbar^2}{2m_\p{r}} \frac{m^2}{\rho^2} \nonumber \\
&+ \tfrac{1}{2}m_\p{r} ( \omega_\rho^2\rho^2 + \omega_z^2 z^2) \, .
\label{eq:h}
\end{align}
Here, 
$\rho$ and $z$ are the relative radial and axial coordinates, respectively. Also, $\omega_{\rho, \p{z}} = 2\pi f_{\rho,\mathrm{z}}$, where $f_{\rho, \p{z}}$ are the corresponding trapping frequencies.
The intermolecular separation is given by $r = \sqrt{\rho^2 + z^2}$, and the projection of the relative motion along the axis of symmetry is given by $m$.
We impose a repulsive short-range potential, $C_{12}/r^{12}$, setting the $C_{12}$ coefficient such that the potential $C_{12}/r^{12} - C_6/r^6$ contains six bound states and gives a scattering length of 100$a_0$.
We neglect the anisotropic $C_6$ coefficient.
While arbitrary, setting the short range part of the potential in this way allows us to conveniently study the important long range effects.

Although the collisions under consideration involve four atoms, the approach described above is justified by the separation in energy scale between the chemical bonds within the ground state dimers ($\sim$THz) and the collision energy or bond between them ($\lesssim\,$MHz).
We also note that the van der Waals coefficient between polar molecules has contributions from the rotation of the molecules as well as the induced dipole moments of the electron clouds. 
An electric field polarizes the molecules and changes the rotational contribution. 
We have calculated the extent of this change and checked that it does not noticeably change the results presented here, as was the case in Ref.~\cite{julienne11}. 
Consequently, we neglect this effect.
Because we confine ourselves to a single collision channel, the resonances we find correspond to shape resonances~\cite{chin_review}, in which the potential experienced by a colliding pair supports a near-degenerate quasi-bound state. We note that Feshbach resonances, in which a colliding pair is coupled to a near-degenerate bound state of a different spin configuration, are possible for the general case of coupling between states of different molecular spin and rotational quantum number.

We first study the two-body energy spectrum.  We solve for eigenstates and
eigenvalues of the Hamiltonian with the potential of \refeq{eq:h}  using
PHAML Version 1.8.0 \cite{phaml, mitchell05}, a parallel two-dimensional finite element
code for elliptic boundary value and eigenvalue problems.  PHAML features
adaptive grid refinement of the discretized spatial coordinates to concentrate
the grid in areas where the wave function varies rapidly, and high order
elements to obtain an accurate solution.  For these computations we used
eighth degree elements.  Within PHAML, ARPACK~\cite{lehoucq98} was used to solve the
discrete eigenvalue problem, using the shift-and-invert spectral transformation
to compute interior eigenvalues, and MUMPS~\cite{amestoy01} to solve the resulting linear
system of equations.  The parallel computations were performed on two nodes of a
Linux cluster.
A particular advantage of the two-dimensional solver is its ability to readily account for the anisotropic interaction and trapping potential. 
By contrast, an expansion in spherical harmonics or non-interacting trap states will struggle to accurately resolve the wavefunction without a very large basis set.
However, for analysis of the wavefunctions we calculate projections onto these functions. 
We solve for the function $F(\rho, z)$, where the full wavefunction is given by $\psi(\rho, z, \phi) = F(\rho, z) e^{im\phi}$. Our bound state calculations consider only $m = 0$, but in the scattering calculations described below we will consider the effects of nonzero $m$.
We study spherically symmetric ($\omega_z = \omega_\rho$) and quasi-2D ($\omega_z \gg \omega_\rho$) geometries, with the dipoles always aligned along the $z-$axis.

\begin{figure}[tb]
\includegraphics[width=0.95\columnwidth, clip]{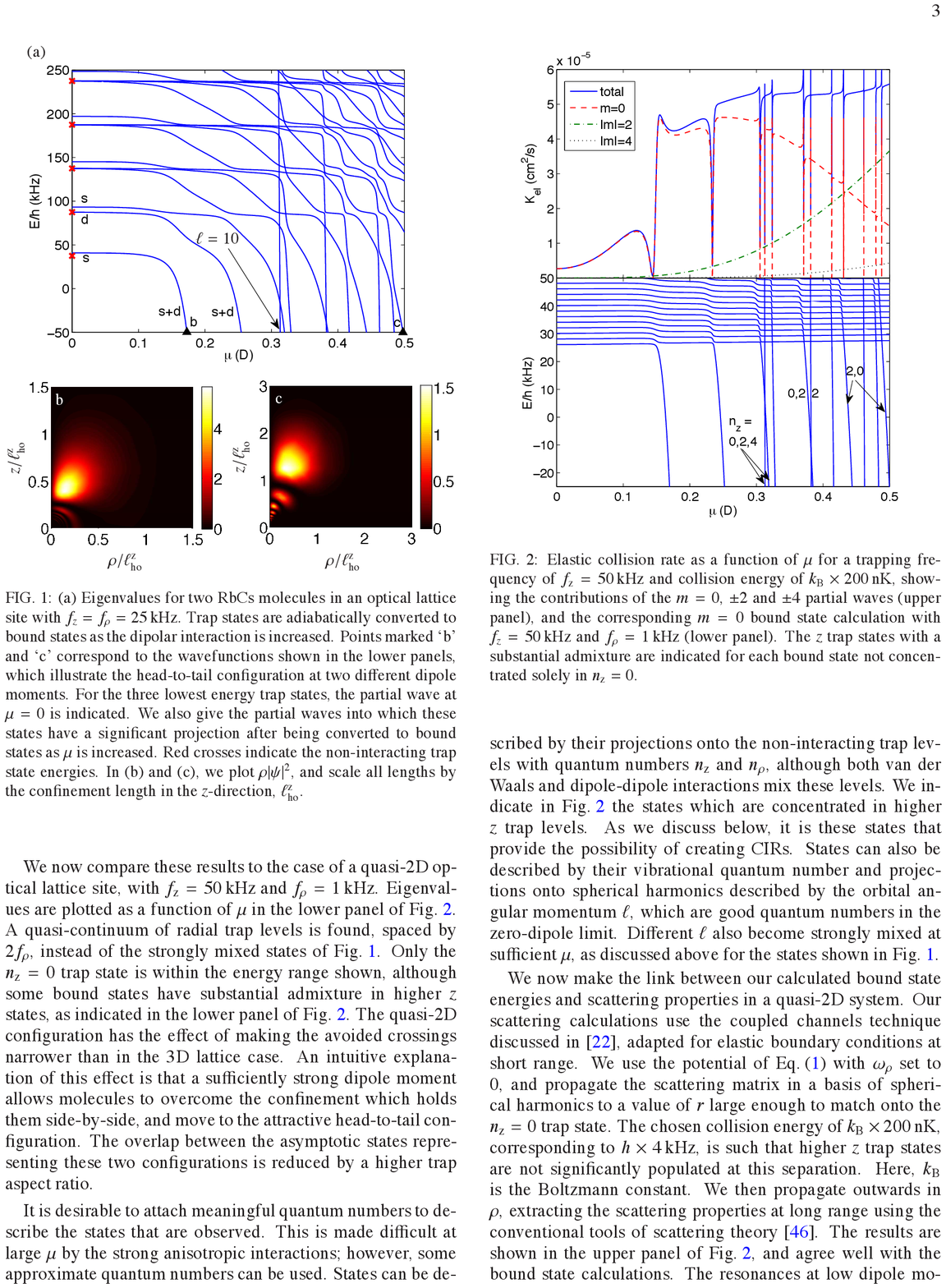}
	\caption{(a) Eigenenergies for two RbCs molecules in an optical lattice site with $f_z = f_\rho = 25\,$kHz. Trap states are adiabatically converted to bound states as the dipolar interaction is increased. Points marked `b' and `c' correspond to the wavefunctions shown in the lower panels, which illustrate the head-to-tail configuration at two different dipole moments. For the three lowest energy trap states, the partial wave at $\mu = 0$ is indicated. We also give the partial waves into which these states have a significant projection after being converted to bound states as $\mu$ is increased. Red crosses indicate the non-interacting trap state energies. 
	In (b) and (c), we plot $\rho |\psi|^2$, and scale all lengths by the confinement length in the $z$-direction, $\ell_\p{ho}^\p{z}$.}
	\label{fig:bnd_d}
\end{figure}
The eigenenergies of two RbCs molecules in a spherically symmetric lattice site with $f_z = f_\rho = 25\,$kHz are shown as  a function of dipole moment in \reffig{fig:bnd_d}a.
We use the term bound states to refer to those that are bound when the trap is adiabatically turned off.
States close to the non-interacting trap level energies, $(n_\p{z} + 1/2)\hbar\omega_\p{z} + (2n_\rho + |m| + 1)\hbar \omega_\rho$, are called trap states.
Here, Bose symmetry allows the trap state quantum numbers to be $n_{\p{z}} = 0,2,4\ldots$, $n_\rho = 0,1,2,\ldots$ and $m = 0, \pm2, \pm4,\ldots$.
At zero dipole moment, the trap state energies are affected by the van der Waals interactions.
Dipole moments above approximately 0.1\,D cause a significant change to these energies.
A large number of avoided crossings occur as trap states are brought into the potential by the increasing dipolar attraction.
All crossings are avoided, although some are too narrow to be visible on the scale shown.
States of different $\ell$ are mixed by the dipolar interactions, with the broadest crossings occuring between states of low $\ell$. 
For example, the first two trap levels converted to bound states as $\mu$ is increased are primarily of mixed $s$- and $d$-wave symmetry, whereas the steeply descending state with a series of narrow crossings near $\mu = 0.31\,$D is concentrated in $\ell = 10$.
Figures \ref{fig:bnd_d}b and \ref{fig:bnd_d}c illustrate the wavefunction near $E/h = -50\,$kHz for dipole moments of $\mu = 0.17\,$D and $\mu = 0.498\,$D, respectively.
We plot the function $\rho |\psi|^2$, to more clearly illustrate both short and large length scales.
For the state at $\mu = 0.17\,$D, vibrational nodes of the bound states of the van der Waals potential can be seen as rings of constant $r < 0.25 \ell_\p{ho}^\p{z}$. A $d$-wave component at long range provides the head-to-tail configuration, with the wavefunction concentrated in the region $\rho < |z|$. This makes $E$ decrease as $\mu$ increases.
The state at $\mu = 0.498\,$D is much more strongly coupled between partial waves, and has a correspondingly more complicated configuration.

\begin{figure}[tb]
\includegraphics[width=0.95\columnwidth, clip]{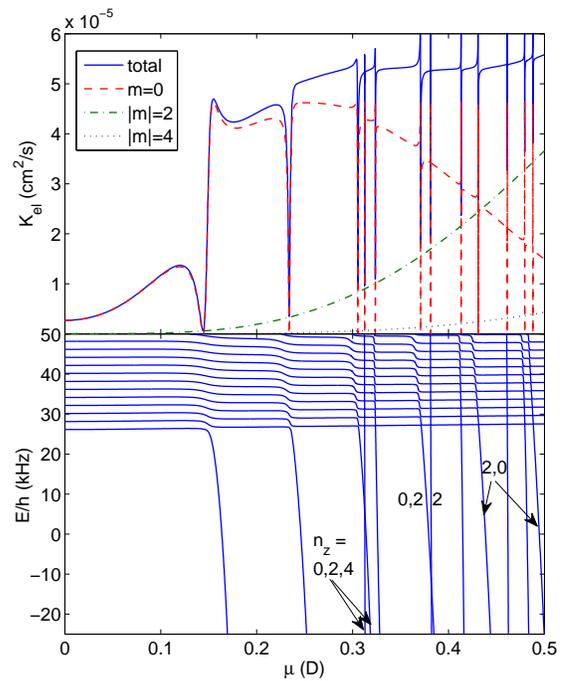}
	\caption{Elastic collision rate as a function of $\mu$ for a trapping frequency of $f_\p{z} = 50$\,kHz and collision energy of $k_\p{B} \times 200$\,nK, showing the contributions of the $m = 0$, $\pm 2$ and $\pm 4$ partial waves (upper panel), and the corresponding $m = 0$ bound state calculation with $f_z = 50\,$kHz and $f_\rho = 1\,$kHz (lower panel). 
	The $z$ trap states with a substantial admixture are indicated for each bound state not concentrated solely in $n_\p{z} = 0$.
}
	\label{fig:2d}
\end{figure}	
We now compare these results to the case of a quasi-2D optical lattice site, with $f_\p{z} = 50\,$kHz and $f_\rho = 1\,$kHz.
Eigenvalues are plotted as a function of $\mu$ in the lower panel of \reffig{fig:2d}.
A quasi-continuum of radial trap levels is found, spaced by $2f_\rho$, instead of the strongly mixed states of \reffig{fig:bnd_d}.
Only the $n_\p{z} = 0$ trap state is within the energy range shown, although some bound states have substantial admixture in higher $z$ states, as indicated in the lower panel of \reffig{fig:2d}.
The quasi-2D configuration has the effect of making the avoided crossings narrower than in the 3D lattice case. 
An intuitive explanation of this effect is that a sufficiently strong dipole moment allows molecules to overcome the confinement which holds them side-by-side, and move to the attractive head-to-tail configuration. 
The overlap between the asymptotic states representing these two configurations is reduced by a higher trap aspect ratio.

It is desirable to attach meaningful quantum numbers to describe the states that are observed. 
This is made difficult at large $\mu$ by the strong anisotropic interactions; however, some approximate quantum numbers can be used.
States can be described by their projections onto the non-interacting trap levels with quantum numbers $n_\p{z}$ and $n_\rho$, although both van der Waals and dipole-dipole interactions mix these levels. 
We indicate in \reffig{fig:2d} the states which are concentrated in higher $z$ trap levels. 
As we discuss below, it is these states that provide the possibility of creating CIRs.
States can also be described by their vibrational quantum number and projections onto spherical harmonics described by the orbital angular momentum $\ell$, which are good quantum numbers in the zero-dipole limit.
Different $\ell$ also become strongly mixed at sufficient $\mu$, as discussed above for the states shown in \reffig{fig:bnd_d}.

We now make the link between our calculated bound state energies and scattering properties in a quasi-2D system.
Our scattering calculations use the coupled channels technique discussed in~\cite{julienne11}, adapted for elastic boundary conditions at short range. 
We use the potential of \refeq{eq:h} with $\omega_\rho$ set to 0, and propagate the scattering matrix in a basis of spherical harmonics to a value of $r$ large enough to match onto the $n_\p{z} = 0$ trap state. 
The chosen collision energy of $k_\p{B} \times 200$\,nK, corresponding to $h \times 4$\,kHz, is such that higher $z$ trap states are not significantly populated at this separation. Here, $k_\p{B}$ is the Boltzmann constant.
We then propagate outwards in $\rho$, extracting the scattering properties at long range using the conventional tools of scattering theory~\cite{taylor72}.
The results are shown in the upper panel of \reffig{fig:2d}, and agree well with the bound state calculations. 
The resonances at low dipole moment are widest and most isolated from other features, making them the most useful for resonant control.
Contributions to the elastic collision rate coefficient from collisions with higher $m$ make the resonance minima nonzero.

\begin{figure}[tb]
	\centering
	\includegraphics[width=0.95\columnwidth, clip]{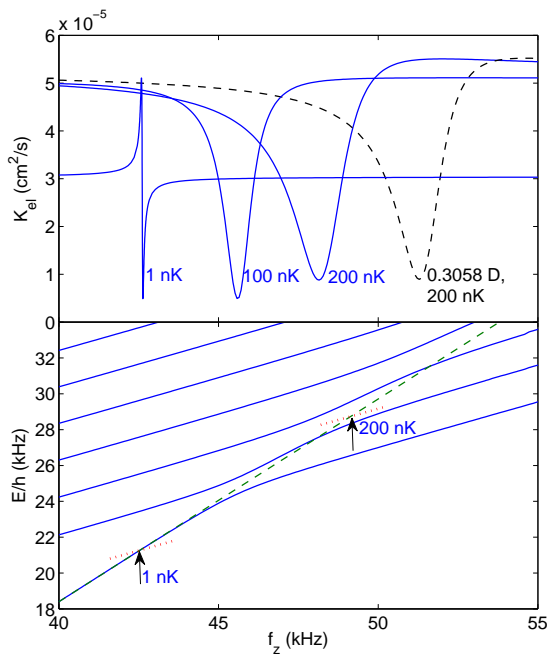}\\	
	\caption{Confinement induced resonances created by tuning of $f_\p{z}$. 
	The top panel shows the elastic collision rate for a quasi-2D trap with $f_\p{z} = 50$\,kHz. 
	We have summed the contributions of partial waves from $m = 0$ to $|m|=4$. 
	Solid lines correspond to $\mu = 0.3048\,$D and collision energies of $k_\p{B} \times 1\,$nK, $k_\p{B} \times 100\,$nK and $k_\p{B} \times 200\,$nK, as labelled. 
	The dashed line ($\mu = 0.3058\,$D) shows the sensitivity of the resonance location to electric field variation. 
The bottom panel shows the eigenenergies of the Schr\"odinger equation with the potential of \refeq{eq:h}, using $\mu = 0.3048$\,D and trapping frequencies of $f_\mathrm{z} = 50$\,kHz and $f_\rho = 1$\,kHz. 
The dashed green line shows the perturbative result of \refeq{eq:pert}. 
Dotted red lines and arrows show the intersections of this calculation with scattering states of the given kinetic energies. These intersections agree well with the calculated resonance locations of the upper panel.
	}
	\label{fig:cir}
\end{figure}
Optical lattices have been used to control scattering lengths in neutral gases~\cite{haller10, froehlich11} by making a state corresponding to an excited trap level near degenerate with the colliding atoms. 
This depends on the scattering length being of the same order as the characteristic length scale of the confinement. 
With the large dipole length characteristic of interactions between polar molecules, it is reasonable to expect that similar confinement induced effects should occur. 
In the lower panel of \reffig{fig:cir} we show this effect for RbCs molecules, calculating eigenenergies as a function of $f_\p{z}$ while maintaining $f_\rho = 1$\,kHz.
We assume $\mu = 0.3048$\,D, corresponding to an easily accessible electric field of 0.67\,kV/cm.
The figure shows six trap levels with $n_\p{z} = 0$ and a range of $n_\rho$, and a single bound state crossing these levels.
As shown in \reffig{fig:2d}, this bound state has substantial admixture in higher $z$ trap states.
The energy of the bound state therefore increases with $f_\p{z}$ faster than the trap levels.

We have perturbatively calculated the change $\delta E$ in an eigenenergy from changing $\omega_\p{z}$ to $\omega_\p{z} + \delta \omega_\p{z}$, which results in
\begin{align}
\delta E = \frac{1}{2} \left\langle \left(\frac{z}{\ell_\p{ho}^\p{z}}\right)^2 \right\rangle \hbar \delta \omega_\p{z}\, .
\label{eq:pert}
\end{align}
Here $\langle \cdots \rangle$ indicates calculating the expectation value with respect to the selected wavefunction.
The result for the bound state of \reffig{fig:cir}, evaluated with the numerically obtained wavefunction at $f_\p{z} = 40\,$kHz, is shown as a green dashed line. 
The red dotted lines correspond to the energy of a non-interacting pair of molecules in the ground trap state, with $k_\p{B} \times 1$\,nK and $k_\p{B} \times 200$\,nK of relative kinetic energy.
The points at which these cross the perturbative calculation correspond well with the scattering resonances shown in the upper panel of Fig.~\ref{fig:cir}. 
For a relative kinetic energy of $k_\p{B} \times 200$\,nK, the feature has a width with respect to $f_\p{z}$ variation of approximately 5\,kHz, although accurate electric field control will be necessary due to the resonance location being strongly dependent on the dipole moment.
This is shown by the dashed line in the top panel, for a dipole moment of 0.3058\,D, which corresponds to a change in the applied electric field of approximately 2.5\,V/cm. 
The temperature is also of significance, as shown by the calculations for collision energies of $k_\p{B} \times 100$\,nK and $k_\p{B} \times 1$\,nK, which produce narrower peaks at lower trapping frequencies.
This result illustrates that the location and properties of the resonances can be controlled by manipulating both the electric field and the confinement.

In conclusion, we have studied the role that an optical lattice can play in controlling the collisional properties of nonreactive polar molecules.  
We have shown that tight confinement allows for much broader avoided crossings, giving a greater resonance width than is available in free space. 
We have also shown that confinement induced resonances can be easily created, with the caveat that their location is sensitive to the dipole moment.
Measurements of resonance locations would constrain the short range potential, for which we studied just one example with a scattering length of 100\,$a_0$.
However, this should not significantly alter the density of states or our finding that RbCs will have several accessible resonances for dipole moments less than 0.5\,D and trapping frequencies on the order of tens of kHz.
These results will therefore be of significance for upcoming experiments using non-reactive polar molecules.

We acknowledge funding from an AFOSR MURI on ultracold molecules (T.M.H. and P.S.J.) and partial funding from the ONR (P.S.J.). We thank Z. Idziaszek for stimulating discussions.

\bibliography{tomsrefs}
\end{document}